\documentclass{llncs}

\ifx\pdftexversion\undefined
  \usepackage[dvips,usenames]{color}
\else
  \usepackage[pdftex,usenames,dvipsnames]{color}
\fi
\usepackage{xspace}
\usepackage[T1]{fontenc}  % access \textquotedbl
\usepackage{textcomp}     % access \textquotesingle
\usepackage{amsmath}
\usepackage{amssymb}
\usepackage{graphics}
\usepackage[utf8]{inputenc}
\usepackage{listings}
\usepackage{cite}  % orders unordered citations, and introduces intervals
\usepackage{hyperref}

\DeclareMathSymbol{\Sigma}{\mathalpha}{operators}{6}  % straight, rather than italics

\definecolor{lightgray}{rgb}{0.97, 0.97, 0.97}
\lstdefinelanguage{minizinc}
{
morekeywords={
ann, annotation, any, array, assert, bool, constraint, else, endif, enum, float, forall, function,
if, in, include, int, list, of, op, output, par, predicate, record, set,
solve, string, test, then, tuple, type, var, where,
abort, abs, acosh, array_intersect, array_union,
array1d, array2d, array3d, array4d, array5d, array6d, asin, assert, atan, bool2int, card,
ceil, combinator, concat, cos, cosh, dom, dom_array, dom_size, dominance,
fix, exp, floor, index_set, index_set_1of2,
index_set_2of2, index_set_1of3, index_set_2of3, index_set_3of3, int2float, is_fixed,
join, lb, lb_array, length, let, ln, log, log2, log10, min, max, pow, product, round, set2array,
show, show_int, show_float, sin, sinh, sqrt, sum, tan, tanh, trace, ub, and ub_array,
minisearch, search, while, repeat, next, commit, print, post, sol, scope, time_limit, break, fail
},
sensitive=false, % are the keywords case sensitive
morecomment=[l][\em\color{ForestGreen}]{\%},
morestring=[b]",
}

\newcommand{\ignore}[1]{}

\newcommand{\gecode}{\textsc{Gecode}\xspace}
\newcommand{\gecopen}{\textsc{Gecode}+$\mathbb{S}$\xspace}
\newcommand{\chuffed}{\textsc{Chuffed}}
\newcommand{\izplus}{\textsc{iZplus}}
\newcommand{\sushi}{\textsc{Sushi}\xspace}
\newcommand{\kaluza}{\textsc{Kaluza}\xspace}
\newcommand{\norn}{\textsc{Norn}\xspace}
\newcommand{\hampi}{\textsc{Hampi}\xspace}
\newcommand{\zstr}{\textsc{Z3-str2}\xspace}
\newcommand{\ascii}{\mathsf{ASC}}
\newcommand{\atoi}{\mathcal{I}}
\newcommand{\maxlen}{\ell}
\newcommand{\str}[1]{\text{\textquotedbl}#1\text{\textquotedbl}}
\newcommand{\n}{\mathbb{N}}
\newcommand{\gcc}{\mathcal{GCC}}
\newcommand{\dfa}{\mathcal{L}_\textnormal{D}}
\newcommand{\nfa}{\mathcal{L}_\textnormal{N}}
\newcommand{\flatstr}{\mathcal{F}^{\textnormal{str}}}
\newcommand{\flatint}{\mathcal{F}^{\textnormal{int}}}
\newcommand{\rew}{\mapsto}
\newcommand{\arr}{\mathcal{A}}
\newcommand{\dom}{\mathcal{D}}
\newcommand{\inn}{~\overline{\in}~}
\newcommand{\var}{\mathcal{V}}
\newcommand{\vararr}{\var_{\text{arr}}}
\newcommand{\varint}{\var_{\text{int}}}
\newcommand{\varstr}{\var_{\text{str}}}

\begin{document}
\title{MiniZinc with Strings}

\author{
  Roberto Amadini  \inst{1} \and 
  Pierre Flener    \inst{2} \and
  Justin Pearson   \inst{2} \and \\
  Joseph D. Scott  \inst{2} \and
  Peter J. Stuckey \inst{1} \and 
  Guido Tack       \inst{3}
}

\institute{
  University of Melbourne, Victoria, Australia \and
  Uppsala University, Uppsala, Sweden         \and
  % omitted for symmetry with the other addresses: Faculty of IT,
  Monash University, Australia
}
\maketitle

\begin{abstract}
Strings are extensively used in modern programming languages and
constraints over strings of unknown length occur in a wide range of real-world 
applications such as software analysis and verification, testing, model checking,
and web security. Nevertheless, practically no CP solver natively supports 
string constraints. We introduce string variables and a suitable 
set of string constraints as builtin features of the MiniZinc modelling language.
Furthermore, we define an interpreter for converting a MiniZinc model with
strings into a FlatZinc instance relying on only integer variables.
This provides a user-friendly interface for modelling
combinatorial problems with strings, and enables both string and non-string
solvers to actually solve such problems. 
\end{abstract}

\setcounter{footnote}{0}  % Springer uses footnote numbers to index institutions

\section{Introduction}
\label{sec:intro}

Strings are widely adopted in modern programming languages for representing 
input/output data as well as actual commands to be executed dynamically. 
The latter is particularly critical for security reasons: consider, e.g., the 
dynamic execution of a malicious SQL query.
% : consider for example the execution of a 
% malicious SQL query that might dump a database or delete entire tables thereof. 
% Apart from security issues, tracking (an approximation of) the 
% possible values of a string variable can also help in bug detection and code 
% optimisation. 
The interest in string analysis---needed in real-life applications
such as test-case generation~\cite{dyn_test_db}, program analysis~\cite{path_feas}, 
model checking~\cite{mod_check}, web security~\cite{waptec}---is active and growing 
\cite{DBLP:conf/aplas/KimCPKR14,DBLP:conf/cc/MadsenA14,DBLP:journals/spe/CostantiniFC15},
and inevitably
implies the processing of string constraints such as string (in-)equality, 
concatenation, and so on. Nevertheless, in the constraint 
programming (CP) context, practically no solver natively supports 
string constraints.
To our knowledge, the only exception is a new 
extension~\cite{ASTRA:PhD:Scott} 
with bounded-length string variables
of \gecode solver~\cite{gecode},
here called \gecopen for convenience,
agreed to become official part of \gecode.
Empirical results show that
% , despite being a prototype,
\gecopen usually is better than dedicated string solvers 
like \hampi~\cite{hampi12}, \kaluza~\cite{kaluza},
and \sushi~\cite{sushi}. %(SMT solvers)
%as well as \sushi~\cite{sushi} (an automaton-based solver).

We take a further step towards the definition and solving of string 
constraints.
% in CP.  --> Pierre: no, that's already done by us in Uppsala!
Our first contribution is the extension of the 
\textit{Mini\-Zinc}~\cite{minizinc} modelling language
by string variables of unknown length. 
MiniZinc enables the specification of constraint problems over 
(sets of) integers and real numbers, but currently does not allow models to contain 
string variables. Thanks to the extension we describe,
a MiniZinc user can now naturally define 
and solve a MiniZinc model containing string variables and 
constraints, as well as other constraints on other types.

Our second contribution concerns the solving of MiniZinc models 
with strings.
% Indeed, as previously stated, there do not currently
% exist mature CP solvers over strings.
Since MiniZinc is designed to interface easily to back-end 
solvers---through the conversion into specialised FlatZinc instances---we 
define suitable redefinitions for converting a MiniZinc model with strings 
into an equivalent FlatZinc instance containing only integer variables. 
This is achieved by bounding the maximum length of a string variable. In this way, 
every solver supporting FlatZinc can now solve a MiniZinc model with strings 
without any manual intervention.
%\textcolor{blue}{
This approach follows the open-sequence representation 
of~\cite{bound_str}. However, we underline that our contribution is orthogonal 
to \cite{bound_str} and 
generalises its work (see Section \ref{sec:flatzinc}). 
The MiniZinc formulation we propose does not impose 
restrictions on the string length (enabling us to express unbounded-length strings) 
and allows any solver to use the preferred string representation (e.g., bit 
vectors, automata, or SMT formulae). Furthermore, we handle a superset of the 
constraints of \cite{bound_str}.
%}

Our third contribution is an experimental evaluation of \gecopen and state-of-the-art 
CP solvers (Chuffed, Gecode, and iZplus) on the \norn string benchmark~\cite{norn}: 
native support for string variables usually pays off, but not always,
in which case the technology of the best solver varies.
%\textcolor{blue}{
Indeed, we prove that---despite longer flattening times---sometimes our 
decomposition can be more beneficial than using a dedicated string solver.
%}

\textit{Paper Structure.} Section~\ref{sec:background} gives the background notions. 
% about string variables, MiniZinc and FlatZinc. 
Sections~\ref{sec:minizinc} and~\ref{sec:flatzinc}
describe the string extensions we implemented for MiniZinc and FlatZinc. 
% (and in 
% particular the conversion from MiniZinc with strings to FlatZinc without strings). 
Section~\ref{sec:evaluation} presents the experimental results before we 
conclude in Section~\ref{sec:concl}.

\section{Background}
\label{sec:background}

\emph{MiniZinc}~\cite{minizinc} is a flexible and user-friendly modelling language 
for representing constraint problems. The motto is 
``\emph{model once, solve anywhere}'': each MiniZinc model is 
solver-independent, although it may 
contain annotations to communicate with the underlying solver. MiniZinc 
supports the most common global constraints (i.e., constraints defined over 
an arbitrary number of variables~\cite{globs_cat}) and allows 
the separation between model and data: a MiniZinc model 
can be defined as a generic template to be instantiated by different data. 
% For 
% example, the $n$-queens problem can be defined in terms of 
% an unspecified number $n$ of queens, and instantiated by
% providing the value of parameter~$n$.

\emph{FlatZinc} is a solver-specific target language for MiniZinc. 
Each MiniZinc model (together with corresponding data, if any) is converted 
into FlatZinc in the form required by a solver. In other terms, from the same 
MiniZinc model different FlatZinc instances can be derived according to 
solver-specific redefinitions.
% For example, if the $n$-queens 
% problem is defined in terms of the $\mathtt{alldifferent}([x_1, \dots, x_n])$ 
% global constraint, a solver can decide to keep the constraint as is or 
% to unfold it into the logical conjunction  
% $\bigwedge_{1 \leq i<j\leq n} x_i \neq x_j$.

Following the approach
of~\cite{hampi12,kaluza,bound_str,ASTRA:PhD:Scott} we focus
in this work on constraint solving over \emph{bounded} string variables, i.e., 
string variables $x$ having a bounded length~$\maxlen$, with $|x| \leq \maxlen \in \n$.
There also are solvers for string variables having \emph{fixed} length 
(e.g., the initial version~\cite{hampi09} of \hampi) or \emph{unbounded} length 
(e.g., \norn~\cite{norn}, \sushi~\cite{sushi}, and \zstr~\cite{z3str2}).
We point out that our MiniZinc
% implementation --> Pierre: really?!?
language extension
allows us to express problems with unbounded string variables.
Note that, while problems over fixed-length string variables are trivially decidable, 
satisfiability with unbounded strings is not decidable in general~\cite{str_decid}.

\emph{Notation}.
Given a fixed alphabet $\Sigma$, a string $x \in \Sigma^*$ is a 
finite sequence of $|x| \geq 0$ characters of $\Sigma$.
Let $\ascii$ denote the set of the ASCII symbols, and define the function 
$\atoi \colon \ascii \to [1, 128]$ such that $\atoi(a) = k \leftrightarrow a$ is 
the $k$-th ASCII symbol.
The symbols $=$, $\neq$, and $\preceq$ indicate respectively string equality, 
inequality, and %non-decreasing 
lexicographical order on $\Sigma^*$.
The concatenation of $x$ and $y$ is 
indicated with $x \cdot y$, while $x^n$ is the iterated concatenation of~$x$
for $n$ times; $x^0$ is the empty string~$\epsilon$, while by $x^{-1}$ is
the reverse of~$x$.
If~$x$ is a string (resp., an array) we indicate with $x[i]$ its $i$-th 
character (resp., element). Indices start from 1 in both cases.
% and with $x[i..j]$ the sub-string 
% $x[i] x[i+1] \dots x[j]$. In both cases, we assume the indices 
% start from 1. 
The symbol $\in$ is used for both  
set membership and character occurrence in a string.
% We use the letters $a, b, c, \dots$ for characters; $l, m, n, \dots$ for integer 
% variables; $x, y, z, \dots$ for string variables; $R, S, T, \dots$ for set of 
% strings; $X, Y, Z, \dots$ for arrays of integer variables.
% \todo{Review this paragraph after writing Section 3.}

\section{MiniZinc with Strings}
\label{sec:minizinc}

MiniZinc supports plenty of builtins (e.g., comparisons, basic and advanced 
numeric  operations, set operations, logical operators, \dots) and 
global constraints.
It currently permits four types of variables (i.e., 
Booleans, integers, floats, and sets of integers) while strings can only be 
fixed literals, 
used for formatting output or defining model annotations.
% 
% and global constraints. It 
% allows comparisons ($=$, $\neq$, $<$, $>$, $\leq$, $\geq$), basic and advanced 
% numeric  operations (e.g., $+$, $-$, $\times$, $/$, $\sqrt{}$, $\exp$, $\log$, $\sin$, $\cos$, 
% \dots), set operations (e.g., $\cup$, $\cap$, $\setminus$, $\subseteq$, \dots), 
% logical operators ($\wedge$, $\vee$, $\neg$, $\rightarrow$, $\leftrightarrow$, $\forall$, 
% $\exists$, \dots), as well as the most common global constraints 
% (\texttt{alldifferent}, \texttt{count}, \texttt{member}, \texttt{regular}, 
% \dots).
% MiniZinc currently permits four types of variables: Booleans, integers, 
% floats, and sets of integers.
% 
% Currently strings in MiniZinc can only be fixed literals, 
% used for formatting output or defining model annotations.

Our first contribution is introducing \textit{string variables}, i.e., 
variables $x \in \Sigma^*$, where $\Sigma$ is a given alphabet. As a first step, 
we assume that the alphabet $\Sigma$ is always the set $\ascii$ of ASCII characters. 
Although we focus on bounded-length strings, we do not impose any limitation on the 
maximum string length~$\maxlen$. 
%We did not introduce a new type for characters.
% see a character $a \in \ascii$ as the corresponding string 
% $a \in \ascii^*$ of length one. 
% \pjs{Make it clear a character is an integer!}
% I wanted to say that we use {``a'', ``b'', ``c''} instead of {'a', 'b', 'c'}. 
% Anyway I agree is not clear, I rephrased.

Figure \ref{ex:var_decl} shows three 
string variable declarations in a MiniZinc model. 
Variable \texttt{x} belongs to $\ascii^*$ but its maximum length is not specified: 
a solver can choose the preferred upper bound~$\maxlen$ for its length or 
consider it unbounded.
For example, a solver using automata for representing strings does not 
need to set a maximum length since it can represent strings of 
arbitrary length.
Conversely, a bounded-length string solver such as \gecopen has to fix a 
maximum string length~$\maxlen$.\footnote{This tricky part is analogous to a 
MiniZinc declaration of the form ``\texttt{var int: x}'' for an integer variable: 
a finite-domain solver assumes the domain \texttt{x} to be finite and chooses
its preferred bounds, while for a MIP solver \texttt{x} is unbounded.}
The length of \texttt{y} can be at most \texttt{N}, 
where \texttt{N} is an integer parameter to be initialised within the model or 
in a separate data file. Variable \texttt{z} even has a constrained alphabet: 
$\mathtt{z} \in \{w \in \{\str{\mathtt{a}}, \str{\mathtt{b}}, 
\str{\mathtt{c}}\}^* \mid |w| \leq 500\}$.
%% wrong:
% = \{\str{\mathtt{a}}, \str{\mathtt{b}}, \str{\mathtt{c}}\}^{500}$.
\begin{figure}[t]
\begin{lstlisting}
  int: N;
  var string: x;
  var string(N): y;
  var string(500) of {"a", "b", "c"}: z;
\end{lstlisting}
\caption{Examples of string variable declarations.}
\label{ex:var_decl}
\end{figure}

Given that we now have string variables, inspired
by~\cite{bound_str,ASTRA:PhD:Scott},
we introduce the string constraints specified in Table~\ref{tab:str_cons}.
\begin{table}[t]
\centering
\caption{MiniZinc string constraints, for each 
$x, y, z \in \ascii^*$, $a, b \in \ascii$, $n, m, q, q_0 \in \n$, $S \subseteq \ascii$, 
$F \subseteq \n$, $D \in \n^{q \times |S|}$, and $N \in \mathcal{P}(\n)^{q \times |S|}$.}
\scalebox{1}{
 \begin{tabular}{|c|c|c|}
 \hline
 Constraint & MiniZinc & Description \\
 \hline
 $x = y,\ x \neq y$ & \texttt{x = y,\  x\,!\!= y} & (in-)equality\\
 $ x \prec y,\ x \preceq y,\ x \succeq y,\ x \succ y $ &
 \texttt{x < y}, \texttt{x <= y}, \texttt{x >= y}, \texttt{x > y} & lexicographic order\\
 $x \in S^*$ & \texttt{x in S} & character set\\
 $x \inn S^*$ & \texttt{str\_alphabet(x, S)} & alphabet \\
 $x \in [a, b]^*$  & \texttt{str\_range(x, a, b)} & character range\\
 $z = x \cdot y$ & \texttt{z = x ++ y} & concatenation\\
 $a = x[n]$ & \texttt{a = x[n]} & character access\\
 $y = x[n..m]$ & \texttt{y = str\_sub(x, n, m)} & sub-string \\
 $y = x^n$ & \texttt{y = str\_pow(x, n)} & iterated concatenation\\
 $y = x^{-1}$ & \texttt{y = str\_rev(x)} & reverse\\
 $n = |x|$ & \texttt{n = str\_len(x)} & length\\
 $x \in \dfa(q, S, D, q_0, F)$ & \texttt{str\_dfa(x, q, S, D, q$_0$, F)} & DFA membership \\
 $x \in \nfa(q, S, N, q_0, F)$ & \texttt{str\_nfa(x, q, S, N, q$_0$, F)} & NFA membership \\
 $\gcc(x, A, X)$ & \texttt{str\_gcc(x, A, X)} & global cardinality \\
 \hline
 \end{tabular}
 }
\label{tab:str_cons}
\end{table}
The constraints $=, \neq, \prec, \preceq, \succeq, \succ$ have the semantics of 
their standard definitions. Let $S \subseteq \ascii$: the semantics of $x \in S^*$ is 
$\forall a: a \in x \to a \in S$, while $x \inn S$ also enforces the 
reverse implication, that is $\forall a: a \in x \leftrightarrow a \in S$.
The constraint 
\texttt{str\_range} offers a shortcut for defining a set of strings over a 
range of characters: $[a, b]^* = \{c \in \ascii \mid a \leq c \leq b \}^*$, so 
for instance $[\str{\mathtt{a}}, \str{\mathtt{d}}]^* = \{
\str{\mathtt{a}}, \str{\mathtt{b}}, \str{\mathtt{c}}, \str{\mathtt{d}}
\}^*$.
%% Already said above:
% The semantics of $x \cdot y$ and $x^n$ is well-known.
%
%Note that the concatenation $z = x \cdot y$ raises length issues: fixed a maximum 
%string length $\maxlen$, we can have that $|z| = |x| + |y| > \maxlen$ even when 
%$|x|, |y| \leq \maxlen$ (analogous considerations hold for $x^n$).
% \pjs{should this be here? I guess implicitly we should assert that $|z| \leq
%   \maxlen$ during solving too!}
%
% RA: True, better to remove.
%
The function $x[i..j]$
% actually  %% Pierre: not sure what the "actually" referred to!
returns 
%\todo{this clashes with the end of Section~2!} RA: I think can be safely removed from Sect. 2}
the substring $x[n] x[n + 1] \dots x[m]$, where $n = \max(1, i)$ and $m = \min(j, |x|)$;
in particular, $i > j$ implies  $x[i..j] = \epsilon$.
%% Already said above:
% While the semantics of $x^{-1}$ and $|x|$ is straightforward,
% The last three constraints of Table~\ref{tab:str_cons} deserve attention.
The constraint $x \in \dfa(q, S, D, q_0, F)$ constrains 
$x$ to be accepted by the deterministic finite automaton (DFA)
$\langle Q, S, \delta, q_0, F \rangle$ where $Q = \{1, \dots, q\}$ is 
the state set, $S = \{a_1, \dots, a_{|S|} \}$ is the 
alphabet, $\delta: Q \times S \to Q$ is the transition function such that 
$D[i, j] = k \leftrightarrow \delta(i, a_j) = k$, $q_0 \in Q$ is the 
initial state, and $F \subseteq Q$ is the set of accepting states. The same applies 
to the non-deterministic finite automaton (NFA) constraint 
$x \in \nfa(q, S, N, q_0, F)$, with the only difference that, while 
$D[i, j] \in Q$, in this case $N[i, j] \subseteq Q$.
Finally, we add a global cardinality constraint 
$\gcc(x, A, X)$ for strings, stating that each character $A[i] \in \ascii$ must occur 
exactly $X[i]$ times in string $x$.

The constraints in Table~\ref{tab:str_cons} express 
all those of~\cite{norn,sushi,hampi12,hampi09,kaluza,z3str2} and 
reflect the most used string operations in modern programming languages. 
We are not aware of string solvers supporting constraints like 
lexicographic ordering and global cardinality, but these are natural for a CP
solver.
Some constraints are redundant, for example since $x[i] = x[i..i]$ and 
$y = x[i..j] \leftrightarrow (\exists y_1, y_2 \in \ascii^*)~ 
x = y_1 \cdot y \cdot y_2 \wedge |y_1| = i - 1 \wedge |y_1 \cdot y| = j$.
The rationale behind such redundancy is to ease the model writing and
to allow solvers to define a specialised treatment for each constraint  
in order to optimise the solving process.

The constraint set we added to MiniZinc is intended to be an 
extensible interface for the definition of string problems to be solved by 
fixed, bounded, and unbounded-length string solvers. 
Consider the MiniZinc model in Figure~\ref{ex:mzn}, encoding 
the problem of finding a minimum-length palindrome string belonging to 
$\{\str{\texttt{a}}, \dots, \str{\texttt{z}}\}^*$, having 
an odd length, and containing the same, positive number of occurrences of
\str{\texttt{a}}, \str{\texttt{b}}, and \str{\texttt{c}}. We can see in this 
example the potential of MiniZinc with strings: the model is succinct and 
readable, it allows the specification of optimisation problems
and not just of satisfaction problems, it 
accepts constraints over different types than just strings, it does not 
impose any bounds on the lengths of the strings, and it enables expressing the 
membership of a string variable to a context-\emph{sensitive} language.

\begin{figure}[t]
\begin{lstlisting}
var int: n;
var string: x;
constraint x = str_rev(x);
constraint str_range(x, "a", "z");
constraint str_len(x) mod 2 = 1;
constraint str_gcc(x, ["a", "b", "c"], [n, n, n]);
constraint n > 0;
solve minimize str_len(x);
\end{lstlisting}
\caption{A model for finding minimum-odd-length palindromes 
with the same, positive number of a's, b's, and c's. 
An optimal solution must have $n = 2 \land |x| = 7$.}
\label{ex:mzn}
\end{figure}
%\pjs{The optimal answer to this is $x = "d"$! I suggest we add
%  \texttt{"abac" = str\_sub(x,i,j)} where i and j are variables?}
% RA: constraint n > 0 ensures str_len(x) in {7, 9, 11, ..}.

\section{FlatZinc with(out) Strings}
\label{sec:flatzinc}

MiniZinc is a solver-independent 
modelling language. In practice, this is achieved by the MiniZinc compiler,
which can translate any MiniZinc model into a specialised FlatZinc instance
for a particular solver, using a solver-specific library of suitable
\emph{redefinitions} for basic and global constraints.

In order to extend MiniZinc with support for string variables, our second
contribution consists of two redefinition libraries to perform different conversions: 
\begin{itemize}
 \item a string-to-string conversion $\flatstr$ that flattens a model $M$ 
 with string constraints into a FlatZinc instance $\flatstr(M)$ with  
 all such constraints preserved;
 \item a string-to-integers conversion $\flatint$ that flattens a model $M$ 
 with string constraints into a FlatZinc instance $\flatint(M)$ with  
 string constraints transformed into integer constraints.
\end{itemize}
The conversion $\flatstr$ is straightforward and we omit its technical details. 
Each string predicate is preserved in the resulting 
FlatZinc instance, with a few exceptions in order to be consistent with
the FlatZinc syntax; e.g., 
% \texttt{str\_alphabet(x, S)} % pjs{Why is a function left behind?} RA: my typo!
% and \texttt{str\_gcc(x, A, X)} are left as is, but 
\texttt{x = y} and \texttt{x\,!\!= y} are rewritten into \texttt{str\_eq(x, y)} and 
\texttt{str\_neq(x, y)} respectively. Similarly, a string function 
is rewritten into a corresponding FlatZinc predicate; e.g., \texttt{n = str\_len(x)} 
is translated into \texttt{str\_len(x, n)}, while \texttt{z = x ++ y} translates into
\texttt{str\_concat(x, y, z)}. This straightforward and fast conversion is 
aimed at solvers supporting (some of) the constraints of 
Table \ref{tab:str_cons}. 
At present, to the best of our knowledge, the only CP solver with such a 
capability is the new \gecopen~\cite{ASTRA:PhD:Scott}.
%% Redundant with end of next paragraph:
% Again, the maximum string length $\maxlen$ is not defined here: the solver decides 
% if and how to bound $\ell$.

When extending MiniZinc with new features, the goal is to be always conservative:
the compiler should produce FlatZinc code executable 
% in the sense that the compiler should be able to translate the new features into
% FlatZinc code that can be run 
by any current FlatZinc-compatible solver, albeit
less efficiently than by a solver with native support for the new features.
Hence we also develop the $\flatint$ conversion.
The underlying idea 
is to map each string variable 
$x$ to an array $X \in [0, 128]^n$ of $n = \min(\overline{|x|}, \maxlen)$
integer variables\footnote{Since $|x|$ can be unknown,
we will use $\overline{|x|}$ here in the translation to
refer to the upper bound on the length of $x$. If $|x|$ is unknown, then the 
notation $(\forall_{i = 1, \dots, |x|})~ P(i)$ actually means 
$(\forall_{i \in [1, \overline{|x|}]})~ i \leq |x| \to P(i)$.}
and an integer variable $\ell_x \in [0, n]$ that represents 
the string length~$|x|$. For $i = 1, \dots, n$ the invariant
$i > \maxlen_x \leftrightarrow X[i] = 0$ enforces that
the end $X[|x| + 1] \dots X[n]$ of the array is padded with trailing zeros.
The main issue here is the maximum size~$\maxlen$, since FlatZinc does not 
allow dynamic-length arrays. We set $\maxlen = 1000$ by default and
issue a warning to the user if an unbounded string variable is
artificially restricted by this transformation.
The user (and in fact each solver) can override this parameter.
%\textcolor{blue}{
The $\flatint$ conversion follows the open-sequence 
representation of \cite{bound_str}. However, we remark that 
this decomposition is only one of the possible choices for solving a 
CP problem with strings, 
implemented here for compatibility with solvers that support FlatZinc and naturally 
handle integer variables.
%}

\begin{figure}[t]
\begin{align}
\varstr(x, n, S) &\rew\, \label{eq_var}
\{\arr(x) \}  \\ 
\arr(x)  &\rew\, 
\langle X \rangle\left.
\begin{cases}
    n = \min(\overline{|x|}, \maxlen),~ \vararr(X, n, 0..\atoi(\dom(x))),\\
    \varint(\maxlen_x, 0..n), \maxlen_x = |x|,
    (\forall_{i \in [1,n]})~ i > \maxlen_x \leftrightarrow X[i] = 0
\end{cases}\hspace*{-10pt}\right\} \label{eq_arr} \\ 
x = y &\rew\, \{|x| = |y|,~ (\forall_{i \in [1, |x|]})~ \arr(x)[i] = \arr(y)[i]\} \label{eq_eq}\\
x \inn S &\rew\, \left.
\begin{cases}
 (\forall_{i \in [1, \overline{|x|}]})~ \arr(x)[i] \in \atoi(S)\cup\{0\}, \\
(\forall_{i \in \atoi(S)})(\exists_{j \in [1, \overline{|x|}]})~ \arr(x)[j] = i
\end{cases} \hspace*{-10pt}\right\} \label{eq_in}\\
%x \neq y &\rew\, \{|x| = |y| \to (\exists_{i \in [1, |x|]})~ \arr(x)[i] \neq \arr(y)[i]\}\\
x \preceq y &\rew\, \{\mathtt{lex\_lesseq}(\arr(x), \arr(y)) \}\label{eq_prec}\\
%\gcc(x, A, X) &\rew\, \{\mathtt{global\_cardinality}(\arr(x), [I(a) ~|~a \in A], X) \}\\
x \cdot y &\rew\, \langle z\rangle\left.
\begin{cases} \varstr(z), |z| = |x| + |y|,
 (\forall_{i \in [1, |x|]})~\arr(z)[i] = \arr(x)[i], \\
 (\forall_{j \in [1, |y|]})~\arr(z)[j + |x|] = \arr(y)[j]
\end{cases}\hspace*{-10pt}\right\} \label{eq_cat}\\
x[i..j] &\rew\, \langle y\rangle\left.
\begin{cases}  n = \max(1, i),~ m = \min(|x|, j),\\
                \varstr(y),~ |y| = \max(0, m - n + 1),\\
                (\forall_{k \in [1, {|y|}]})~ \arr(y)[k] = \arr(x)[k + n - 1]
\end{cases}\hspace*{-10pt}\right\} \label{eq_sub}\\
% x^{-1} &\rew\, \langle y\rangle\{|x| = |y|, (\forall_{i \in [1, |x|]})~ 
%               \arr(y)[i] = \arr(x)[|x| - i + 1] \}\\
% |x| &\rew\, \langle y\rangle\{0 \leq n \leq \ell\}\\
% x^n &\rew\, \langle y\rangle\{|y| = n|x|, (\forall_{i \in [1, |x|], j \in [1, |y|]}) 
%             \arr(y)[|x|(j - 1) + i] = \arr(x)[i] \}\\
%x \in S &\rew\, \{(\forall_{i \in [1, |x|]})~ \arr(x)[i] \in \atoi(S) \}\\
x \in \dfa(q, S, &D, q_0, F) \rew\,\label{eq_dfa} \nonumber\\
&\left.\begin{cases} 
s = |S| + 1,~ D' \in [1, q]^{q \times s},~ T = \mathtt{sort}(\atoi(S)), \\
(\forall_{i \in [1, q], j \in [1, s]})~ D'[i, j] = \begin{cases}
            0 & \text{if $j = 1 \wedge D[i, j] \notin F$} \\
            D[i, j] & \text{otherwise}
           \end{cases} \\
\vararr(X, |x|, 0..|x|),~ \mathtt{regular}(X, q, s, D', q_0, F),\\
(\forall_{i \in [1, \overline{|x|}]})~ \arr(x)[i] = \begin{cases}
              T[X[i] - 1] & \text{if $X[i] > 1$}\\
              0 & \text{otherwise}
             \end{cases} 
\end{cases}\hspace*{-10pt}\right\}
\end{align}
\vspace*{-10pt}
\caption{Examples of rewrite rules of $\flatint$.}\label{fig:rules}
\end{figure}

$\flatint$ works through rewrite rules, some of which are listed in
Fig.~\ref{fig:rules}. Each rule has either the form 
$P \rew \{C_1, \dots, C_n\}$, meaning that predicate $P$ is rewritten 
into constraint $C_1 \wedge \dots \wedge C_n$, or the form
$F(x_1, \dots, x_k) \rew \langle E \rangle \{C_1, \dots, C_n\}$, meaning that 
function $F$ is rewritten into expression $E$ subject to 
$C_1 \wedge \dots \wedge C_n$. We use a more readable meta-syntax instead of 
the MiniZinc/FlatZinc one, upon denoting
by $\dom(x) \subseteq \ascii$ the auxiliary function that returns 
the set of characters that may occur in $x$, and by 
$\atoi(S)$ the set $\{\atoi(a) \mid a \in S\}$. Given $D \subseteq \n$ and 
$S \subseteq \ascii$,
the constructs 
$\varint(n, D)$,
$\varstr(x, m, S)$, 
and $\vararr(X, m, D)$ denote respectively  
an integer variable declaration \texttt{var D: n}, 
a string variable declaration \texttt{var string(m) of S: x}, 
and an array of integer variables declaration 
\texttt{array[1..m] of var D: X}.
If a parameter is omitted, then we 
assume $D = [0, 128]$, $m = \ell$, and $S = \ascii$.
% Note that
The entire conversion is specified using MiniZinc itself and
does not require any modifications to the MiniZinc compiler.

Rule \ref{eq_var} of Fig.~\ref{fig:rules} transforms a declaration of 
a string variable $x$ into the 
corresponding declaration of an array $X$ of integer variables via the 
$\arr(x)$ function of Rule \ref{eq_arr}, which
enforces the properties of $X$ described above.
It is important to note that this transformation relies on the
\textit{same} array of integer variables being returned by $\arr(x)$
for a variable $x$ even if the function is called multiple times.
This is achieved through the common subexpression elimination
mechanism built into MiniZinc functions~\cite{stuckey:2013:0:minizincwithfunctions}.
% Note that $X[i]$ is not always fixed, since $i$ can be 
% an unknown: in this case the \texttt{element} global constraint is 
% implicitly used. \pjs{$i$ must be fixed in rule (1)} Yes
% \pjs{Is the iteration in rule (3) $1,|x|$ or $1,\overline{|x|}$ I guess it
%   should be the former?? RA: the latter (even if the former is sound, there is 
%   no need of reification here since also the 0 elements must match).
%   What if $i$ and $j$ are variables? Hmm not sure how
% this rule can work? RA: In (4) I guess? If so I agree.}

Rules \ref{eq_eq} to \ref{eq_prec} are examples of predicate rewriting.
The latter rule takes advantage of 
MiniZinc expressiveness by rewriting $x \preceq y$ in terms of the
\texttt{lex\_lesseq} global constraint over integers. 
Analogously, the $\gcc$ constraint is 
mapped to \texttt{global\_cardinality}. 
Similarly for the $\prec$, $\succ$, $\succeq$, $\neq$, and $\in$ predicates.

Rules \ref{eq_cat} and \ref{eq_sub} 
are examples of function rewriting:
% \pjs{In rule (5) again have to be careful about defining the length of $z$! 
% In (6)? Why?}
% \pjs{In rule (7) is $m = \min(|x|,j)$ or $m = \min(\overline{|x|},j)$ I
%   guess it clear it should be the former! RA: Yes it is.}
a string variable is created, 
constrained, and returned.
The rules for $x^{-1}$, $x^n$, $x[n]$, and $|x|$ are analogous.

Rule \ref{eq_dfa} is tricky. Indeed, the global constraint \texttt{regular}
cannot straightforwardly encode $x \in \dfa(q, S, D, q_0, F)$ since the 
``empty character'' $0$ might occur in $\arr(x)$.
In order to agree with 
the semantics of \texttt{regular}, it is necessary to increment the number $s$ of 
its symbols (so, the $i$-th character of $S$ becomes the $(i + 1)$-st 
symbol of the DFA encoded by \texttt{regular}),
and to add a column at the head of $D$ for dealing with the $0$ character
(matrix $D'$ is the result of this addition---note that the
state $0$ is always a failing state).\footnote{Details:
\url{http://www.minizinc.org/2.0/doc-lib/doc-globals-extensional.html}} 
If \texttt{regular} is satisfiable, then the accepted sequence $X$ is 
re-mapped to a corresponding string thanks to the auxiliary array $T$.

The $\flatint$ converter enables the solving of string problems 
by \emph{any} solver. Clearly, this is achieved at the expense of efficiency. 
Indeed, several new constraints and reifications are introduced.
Consider 
for example the model $M$ of Fig.~\ref{ex:mzn}: the $\flatstr(M)$ conversion 
is instantaneous and produces a FlatZinc instance of only 14 lines, while the 
default $\flatint(M)$, with maximum length $\maxlen = 1000$, is
considerably slower and 
produces 45,011 lines. % about 4 seconds
Obviously the complexity is proportional to $\maxlen$: e.g., 
$\flatint(M)$ consists of 4,511 lines if we set $\maxlen = 100$.

\section{Evaluation}
\label{sec:evaluation}

Our third contribution is an evaluation of our framework on 
\gecopen~\cite{ASTRA:PhD:Scott}
and state-of-the-art CP solvers, namely
\gecode~\cite{gecode} (a finite-domain solver),
%\footnote{We used version 4.4.0. The current development version performs much better due to an improved default search heuristic.} 
\chuffed\cite{chuffed} (based on lazy clause generation~\cite{lcg}), and 
\izplus~\cite{izplus} (which also exploits local search).
There is a lack of standardised and challenging string 
benchmarks~\cite{bound_str,ASTRA:PhD:Scott}.
However, we stress that our goal is \emph{not} an evaluation of solver
performance, but the introduction of a framework to model string problems
easily, for solving by both string and non-string solvers.
%\textcolor{blue}{
Moreover, one of the benefits of introducing strings in MiniZinc 
is the possibility of implementing and comparing challenging 
and standard benchmarks.
%}
% The \sushi and \kaluza benchmarks used in \cite{bound_str} turn out 
% to be trivial to solve. 
We picked five problems of the \norn benchmark \cite{norn}:
$\textnormal{a}^n \textnormal{b}^n$, ChunkSplit, HammingDistance, 
Levenshtein, and StringReplace.
We also used our Palindrome problem of Fig.~\ref{ex:mzn}.
For each problem, we
wrote a MiniZinc model $M$ with parametric bound~$\maxlen$ on string length; 
obtained FlatZinc instances $F_M(f, i)$ by flattening 
$M$ with $f \in \{\flatstr, \flatint\}$ and $\maxlen = i$; and
solved each $F_M(\flatstr, i)$ with \gecopen (we extended the 
FlatZinc interpreter of \gecode for handling $\flatstr$ builtins) and 
each $F_M(\flatint, i)$ with the other solvers, 
upon varying $i \in \{250, 500, 1000\}$.\footnote{
We ran the experiments on Ubuntu machines with 16 GB of RAM and 2.60 GHz 
Intel\textsuperscript{\textregistered} i7 CPU.
The code is publicly available
at \url{https://bitbucket.org/jossco/gecode-string}.}

\begin{table}[t]
 \caption{Runtimes of the solvers. The 
`t/o' abbreviation means that the time-out was reached,
while bold font indicates when a solver performs better than \gecopen.
 }\label{tab:results}
 \centering
 \begin{tabular}{|c|rrr|rrr|rrr|rrr|}
 \cline{2-13}
 \multicolumn{1}{c|}{~} & \multicolumn{3}{c|}{\chuffed} & \multicolumn{3}{c|}{\gecode}  & \multicolumn{3}{c|}{\izplus} & \multicolumn{3}{c|}{\gecopen}\\
  \hline
$\maxlen$ & 250 & 500 & 1000 & 250 & 500 & 1000 & 250 & 500 & 1000 & 250 & 500 & 1000\\
 \hline
$\textnormal{a}^n \textnormal{b}^n$ & 0.9 & \textbf{1.9} & \textbf{4.6} & 2.6 & 16.5 & 132.7 & 1.9 & 6.9 & \textbf{21.8} & 0.4 & 2.7 & 28.1\\
ChunkSplit & 4.9 & 15.6 & t/o & 3.5 & \textbf{8.5} & \textbf{27.0} & 17.5 & 16.3 & \textbf{72.1} & 1.4 & 13.6 & 184.7\\
Hamming & 26.2 & 318 & t/o & 85.7 & t/o & t/o & t/o & t/o & t/o & 0.7 & 5.7 & 56.6\\
Levenshtein & 1.3 & 2.6 & 6.0 & 1.2 & 2.4 & 5.4 & 3.6 & 18.4 & 8.4 & 0.1 & 0.1 & 0.1\\
StringReplace & 2.5 & 7.1 & 24.5 & t/o & t/o & t/o & 3.2 & 7.3 & 39.3 & 0.1 & 0.3 & 1.5\\
Palindrome & \textbf{1.7} & \textbf{24.7} & \textbf{99.0} & t/o & t/o & t/o & \textbf{0.8} & \textbf{2.3} & \textbf{7.0} & n/a & n/a & n/a\\
\hline
 \textit{mzn2fzn} & 70.4 & 55.2 & 48.5 & 68.0 & 51.2 & 34.1 & 58.5 & 49.2 & 46.7 & 17.0 & 11.2 & 10.2\\
 \hline
 \end{tabular}
\end{table}

Table \ref{tab:results} shows the runtimes, in seconds, to conclude the 
search, i.e., the time needed by a solver to prove the (un-)satisfiability of a problem 
(for satisfiability problems) or to find an optimal solution 
(for the only optimisation problem Palindrome). 
We set a solving timeout of $600$ seconds for each problem.
Note that all the runtimes in Table \ref{tab:results} include also the FlatZinc flattening time. The 
\emph{mzn2fzn} row shows the average percentage of the total solving time
 (when a problem is solved) taken for flattening a MiniZinc model.

The $\gcc$ constraint for strings has---to the best of our knowledge---not 
been proposed before in the literature and \gecopen currently does not support it, 
hence its `n/a' time for the palindrome problem.
Our MiniZinc extension (see Table~\ref{tab:results}) covers all the constraints 
implemented by \gecopen.

The message of this evaluation is twofold.
On the one hand, \gecopen is by far the best solver overall, due to its 
native string support and short flattening times to FlatZinc
(see row \textit{mzn2fzn}).
On the other hand, solvers without native string support
sometimes benefit from $\flatint$ for being faster than \gecopen despite longer 
flattening times. This is interesting 
and should stimulate further development of native string support 
in constraint solvers.
%The performance variability observable in Table \ref{tab:results} might 
%also suggest the use of a portfolio solver~\cite{sunnycp}.
% The slowdown of \gecopen compared to \gecode on ChunkSplit for $i>250$
% seems due to its current lack of an $n$-ary concatenation constraint.

\section{Conclusions}
\label{sec:concl}

We present an extension of the MiniZinc language that allows users to model 
and solve combinatorial problems with strings. The framework we 
propose is expressive enough to encode the most used string 
operations in modern programming languages, and---via proper FlatZinc 
translations---it also enables both string and non-string solvers to 
solve such problems. As an example, MIP solvers having a FlatZinc 
interface (e.g., the well-known 
Gurobi~\cite{gurobi} and CPLEX~\cite{cplex}) can now solve string 
problems without manual intervention.

%\textcolor{blue}{
We are not aware of similar work in CP, and
%}
we see our work as a solid starting point for the handling of string 
variables and constraints with the MiniZinc toolchain. We hope our extension encourages 
the development of solvers that can natively deal with strings.
This will hopefully lead to the creation of new, challenging string 
benchmarks, and to the development of dedicated search heuristics 
(e.g., heuristics based on character frequencies in a string).
We are planning to enhance the framework by adding new search annotations, 
constraints, and features, as well extending the string domain from 
ASCII to other alphabets, such as Unicode.  Furthermore,
non-character alphabets could be useful, such as for the
generation of protocol logs~\cite{ASTRA:TAP2015}, where the natural model
would use strings of timestamps.

\bibliographystyle{plain}
\bibliography{biblio}
\end{document}